%% file: main.tex
\documentclass[]{spie}  %>>> use for US letter paper
\usepackage[unicode=true, colorlinks=true, allcolors=blue]{hyperref}

\input{new-macros}

\title{Closing the loop as an inverse problem: the real-time control of Themis adaptive optics}

\author[a]{Éric Thiébaut}
\author[a]{Michel Tallon}
\author[a]{Isabelle Tallon-Bosc}
\author[b]{Bernard Gelly}
\author[b]{Richard Douet}
\author[a]{Maud Langlois}
\author[b]{Gil Moretto}

\affil[a]{Univ. Lyon, Univ. Lyon 1, ENS de Lyon, CNRS, Centre de Recherche Astrophysique de Lyon UMR5574, F-69230, Saint-Genis-Laval, France}
\affil[b]{CNRS IRL20009 FSLAC, 38205 Laguna, Tenerife, Spain}

\authorinfo{Further author information: E-mail: eric.thiebaut@univ-lyon1.fr}

\newcommand*{\estim}[1]{\widehat{#1}}
\newcommand*{\mean}[1]{\overline{#1}}

\newcommand*{\usec}{\micro\second}
\newcommand*{\GHz}{\giga\hertz}

\newcommand*{\Nact}{n_\Tag{act}}
\newcommand*{\Ndat}{n_\Tag{dat}}
\newcommand*{\Nwav}{n_\Tag{wav}}
\newcommand*{\Nsub}{n_\Tag{sub}}
\newcommand*{\noun}[1]{\textsc{#1}\xspace}
\newcommand*{\Themis}{\noun{Themis}}
\newcommand*{\Alpao}{\noun{Alpao}}

\begin{document}
\maketitle

\begin{abstract}
  We have taken advantage of the implementation of an adaptive optics system on
  the \Themis solar telescope to implement innovative strategies based on an
  inverse problem formulation for the control loop.  Such an approach
  encompassing the whole system implies the estimation of the pixel variances
  of the Shack-Hartmann wavefront sensor, a novel real-time method to extract
  the wavefront slopes as well as their associated noise covariance, and the
  computation of pseudo-open loop data.  The optimal commands are computed by
  iteratively solving a regularized inverse problem with spatio-temporal
  constraints including Kolmogorov statistics.  The latency of the dedicated
  real-time control software with conventional CPU is shorter than
  $300\,\micro\second$ from the acquisition of the raw $400×400$ pixel
  wavefront sensor image to the sending of the commands.
\end{abstract}

\keywords{solar adaptive optics, inverse problems framework, real-time control,
  iterative methods, sufficient statistic, pseudo-open loop}

\section{OUTLINE}
\label{sec:outline}

\begin{figure}
  \centering
  \includegraphics[width=\textwidth]{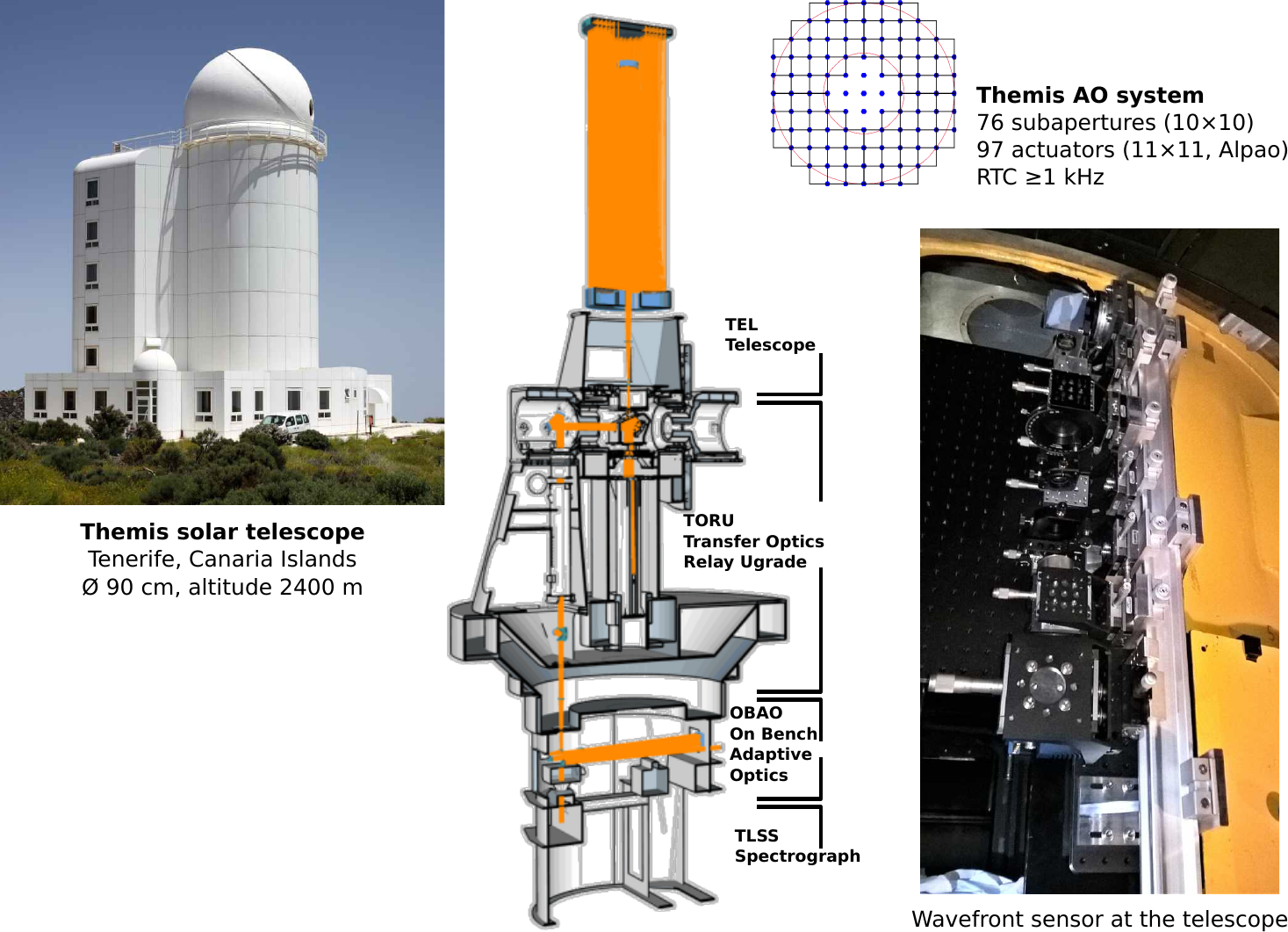}
  \caption{The \Themis AO system.  Left: outside view of the \Themis telescope.
    Center: optical layout of the telescope with the AO system and the
    spectrograph.  Top: layout of the sub-pupils of the Shack-Hartmann
    wavefront sensor and of the actuators of the deformable mirror.  Right: the
    wavefront (camera and optics).}
  \label{fig:Themis-layout}
\end{figure}

The \Themis solar telescope has been equipped with an adaptive optics (AO)
system comprising a 92-actuator \Alpao deformable mirror (DM) and a 10×10
Shack-Hartmann wavefront sensor (SH-WFS) operated at $1\,\kilo\hertz$ (see
Figure~\ref{fig:Themis-layout}).  Developing the real-time control (RTC) of
this AO system was the opportunity to implement innovative strategies studied
in our team for a decade\cite{Thiebaut_Tallon-2010-FRiM, Bechet_et_al-2009-MAP,
  Bechet_et_al-2012-Amsterdam, Thiebaut_et_at-2018-solar_adaptive_optics} and
all based on an inverse problem formulation of the estimation of the commands
applied to the DM.  The inverse problem approach is known to yield excellent
estimators which optimally combine all available information but is usually not
fully used for real-time systems because of the computational burden and of the
amount of required information (e.g., all covariance matrices must be
provided).  Since sufficient statistics are required at all processing stages,
we developed specific detector calibration and pre-processing procedures to
compensate for the non-uniformity of the WFS detector response and optical
throughput and to yield estimated pixel
variances\cite{Tallon_et_al-2022-WFS_SPIE}.  We also developed a novel method
to extract, in real-time, the wavefront slopes and the covariance matrix of
their noise from the images of the WFS despite the very low contrast and the
temporal evolution of the structures at the Sun's
surface\cite{Tallon_et_al-2022-WFS_SPIE}.  For the control, the chosen approach
leads to work with pseudo-open loop data, which requires a good model of the
so-called interaction matrix of the system.  This matrix is calibrated by
measuring the effects of sending stochastic commands
(Section~\ref{sec:calibration}).  The optimal commands are eventually estimated
by iteratively solving a regularized inverse problem
(Section~\ref{sec:control}).  Various spatio-temporal constraints are
implemented including a priori Kolmogorov covariance for the wavefront.  The
dedicated RTC is implemented on a 4-core conventional CPU running under Linux
and is able to close the loop even in harsh conditions with a latency shorter
than $300\,\micro\second$ from the acquisition of the raw $400×400$ pixel WFS
image to the sending of the commands (Section~\ref{sec:RTC}).

\section{OPTIMAL CONTROL}
\label{sec:control}

An adaptive optics (AO) system consists in a wavefront sensor (WFS), a
deformable mirror (DM), and real time control (RTC) hardware and software to
compute and send the commands for the actuators of the DM given the
measurements of the WFS.  In this section, we aim at deriving optimal control
commands for the AO system under clearly specified assumptions.

\subsection{Model of the Wavefront Sensor Data}

At each time frame $t$, the wavefront sensor delivers data $\V d_t \in
\Reals^{\Ndat}$ whose linearized model writes:
\begin{equation}
  \V d_t = \M S ⋅ (\V w_t + \M M ⋅ \V a_t) + \V z_t
  \label{eq:wfs-data-model}
\end{equation}
where $\M S \in \Reals^{\Ndat×\Nwav}$ is the linear response matrix of the
sensor, $\V w_t \in \Reals^{\Nwav}$ is the wavefront during the exposure of
frame $t$, $\M M\in \Reals^{\Nwav×\Nact}$ is the linear response matrix of the
deformable mirror, the so-called \emph{influence matrix}, $\V a_t \in
\Reals^{\Nact}$ is the vector of actuators commands during this frame, and $\V
z_t \in \Reals^{\Ndat}$ accounts the contribution of the noise.  Here $\Ndat$
is the number of measurements per frame, $\Nact$ is the number of actuators,
and $\Nwav$ is the size of the wavefront $\V w_t$ which can be arbitrarily
large even though it is finite in practice.  For a Shack-Hartmann wavefront
sensor, the measurements are the wavefront slopes for each sub-pupil, hence
$\Ndat=2\,\Nsub$ with $\Nsub$ the number of sub-pupils.

In what follows and for the sake of simplicity, we assume that the mirror
linear response $\M M$ and the wavefront sensor linear response $\M S$ are
stable in time.  This can be easily relaxed by adding a subscript $t$ for these
matrices.

\subsection{Optimal Commands}

At any time frame $t$, the best commands are those that will minimize the
residual wavefront aberrations $\V w_{t+δt} + \M M ⋅ \V a_{t+δt}$ at frame
$t+δt$ where $δt > 0$ is the time delay (in number of frames) for the wavefront
sensor to effectively \emph{see} the effects of given actuators commands.  For
example, to maximize the Strehl we would like to apply the following commands:
\begin{equation}
  \V a_{t+δt} = \argmin_{\V a}\Avg[\big]{
    \Norm*{\V w_{t+δt} + \M M ⋅ \V a}^2\Given\V\theta_t}
  \label{eq:least-mse}
\end{equation}
where $\V\theta_t$ represents known information at time $t$ and $\Avg*{\V
  x\Given\V\theta}$ denotes the conditional expectation of $\V x$ knowing
$\V\theta$.  In other words, we want to reduce as much as possible the
residuals on the mean at time $t+δt$ by exploiting all available information at
time $t$.  Thanks to taking the expectation of the residual errors, only the
statistics\footnote{in fact, just the first and second moments}, not the actual
realization, of the future wavefront $\V w_{t+δt}$ need to be known and the
above problem has a closed form solution:
\begin{equation}
  \V a_{t+δt} = -\M M^\dagger⋅\mean{\V w}_{t+δt|t},
  \label{eq:optimal-commands}
\end{equation}
where $\M M^\dagger = \Paren[\big]{\M M\T⋅\M M}^{-1}⋅\M M\T$ is the pseudo
inverse of $\M M$, the linear response of the deformable mirror, and $\mean{\V
  w}_{t+δt|t} = \Avg[\big]{\V w_{t+δt}\Given\V\theta_t}$ is the expectation
of the wavefront at time $t+δt$ knowing information available at time $t$.
This apparently simple solution turns out to be quite challenging to compute in
a real-time system.  We develop in what follows the equations leading to
$\mean{\V w}_{t+δt|t}$ with as few assumptions as possible to remain general.

\subsection{Known Information and Pseudo Open-Loop Data}

At every time frame $t$, new information is brought by the wavefront sensor
data $\V d_{t}$ and by the corresponding commands $\V a_{t}$.  Hence
$\V\theta_{t} = \V\theta_{t-1} \cup \Brace*{\V d_{t},\V a_{t}}$ expresses the
information gain between frames $t-1$ and $t$ and is to be used with
Eq.~\eqref{eq:wfs-data-model}.  It is however equivalent and simpler to
consider the pseudo open-loop\cite{Gilles-2005-closed_loop_stability} (POL)
data instead:
\begin{equation}
  \V y_t = \V d_t - \M G ⋅ \V a_t = \M S ⋅ \V w_t + \V z_t
  \label{eq:pol-data-model}
\end{equation}
where $\M G = \M S⋅\M M$ is the so-called \emph{interaction matrix}.  Then,
the updating rule for the available information is just:
\begin{equation}
  \V\theta_{t} = \V\theta_{t-1} \cup \Brace*{\V y_{t}}.
  \label{eq:update-info}
\end{equation}
Thanks to this change of variables, the commands will not explicitly appear in
the equations leading to $\mean{\V w}_{t+δt|t}$.

\subsection{Updating Rules}

Equation~\eqref{eq:optimal-commands} readily shows that obtaining the best
commands is a matter of being able to predict the expected shape of the
wavefront in a near future.  The wavefronts at different times certainly come
into play in the problem and, to be as general as possible, we consider the
spatio-temporal wavefront $\V x$ which is the concatenation of the wavefronts
in all (future, current, and past) temporal frames:
\begin{equation}
  \begin{array}{lcccccccr}
    \V x = \bigl( & \cdots & \V w_{t-2}\T & \V w_{t-1}\T & \V w_{t}\T & \V w_{t+1}\T & \V w_{t+2}\T & \cdots & \bigr)\T\\
  \end{array}.
  \label{eq:wavefront-sequence}
\end{equation}
From Eq.~\eqref{eq:pol-data-model}, the spatio-temporal wavefront $\V x$ is
related to the pseudo open-loop data $\V y_t$ in the $t$-th frame by:
\begin{equation}
  \V y_t = \M H_t⋅\V x + \V z_t,
  \label{eq:pol-data-model-wrt-x}
\end{equation}
where $\M H_t$, the extended linear model matrix of the data, has the following
block structure with blocks of size $\Ndat×\Nwav$:
\begin{equation}
  \begin{array}{lcccccccr}
    \M H_t = \bigl( & \cdots & \M 0 & \M 0 & \M S & \M 0 & \M 0 & \cdots & \bigr).\\
           & \cdots &  t-2 &  t-1 &   t  &  t+1 & t+2 & \cdots & \\
  \end{array}
  \label{eq:H_t}
\end{equation}

At arrival of data frame $t$, the \emph{maximum a posteriori} (MAP) estimator
of $\V x$ can be expressed as:
\begin{align}
  \V x_{|t}
  &\bydef \argmax_{\V x} \Pr\Paren[\big]{\V x \Given \V\theta_{t}} \\
  &= \argmax_{\V x} \Pr\Paren[\big]{\V x \Given \V y_t, \V\theta_{t-1}}
  &&\text{(since $\V\theta_{t} = \V\theta_{t-1} \cup \Brace*{\V y_{t}}$)}\\
  &= \argmax_{\V x} \Pr\Paren[\big]{\V x, \V y_t \Given \V\theta_{t-1}}
  &&\text{(by Bayes' rule and as $\Pr\Paren[\big]{\V y_t \Given \V\theta_{t-1}}$
    is constant in $\V x$)}\\
  &= \argmax_{\V x} \Pr\Paren[\big]{\V y_t \Given \V x, \V\theta_{t-1}}
  \,\Pr\Paren[\big]{\V x \Given \V\theta_{t-1}}
  &&\text{(again by Bayes' rule)}.
\end{align}
For an AO system, the noise and the wavefronts are independent and the noise
terms in two different frames are mutually independent.  As a consequence, the
pseudo open-loop data $\V y_t$ only depend on the wavefront $\V w_t$ and thus
$\Pr\Paren[\big]{\V y_t \Given \V x, \V\theta_{t-1}} = \Pr\Paren[\big]{\V y_t
  \Given \V x}$.  The conditional expectation and covariance of the pseudo
open-loop data can then be derived as follows:
\begin{align}
  \Avg*{\V y_t \Given \V x}
  &= \Avg*{\M H_t⋅\V x + \V z_t \Given \V x}
  &&\text{(from Eq.~\eqref{eq:pol-data-model-wrt-x})}\notag\\
  &= \M H_t⋅\V x + \Avg*{\V z_t \Given \V x}
  &&\text{(by linearity of the expectation)}\notag\\
  &= \M H_t⋅\V x,
\end{align}
the latter assuming the noise is centered in the sense that $\Avg*{\V z_t
  \Given \V x} = \V 0$\footnote{note that $\Avg*{\V z_t \Given \V x} = \V 0
  \Longrightarrow \Avg*{\V z_t} = \V 0$}, and:
\begin{align}
  \Cov*{\V y_t \Given \V x}
  &= \Cov*{\M H_t⋅\V x + \V z_t \Given \V x}
  &&\text{(from Eq.~\eqref{eq:pol-data-model-wrt-x})}\notag\\
  &= \M H_t⋅\Cov{\V x \Given \V x}⋅\M H_t\T + \Cov{\V z_t \Given \V x}
  &&\text{($\V x$ and $\V z_t$ mutually independent)}\notag\\
  &= \Cov{\V z_t}
  &&\text{(idem and since $\Cov{\V x \Given \V x} = \M 0$)}.
\end{align}
Assuming Gaussian statistics for all variables, then yields:
\begin{align}
  \V x_{|t}
  &= \argmin_{\V x} \Brace*{
    \Norm*{\V y_t - \M H_t⋅\V x}_{\Cov{\V z_t}^{-1}}^2
    + \Norm*{\V x - \mean{\V x}_{|t-1}}_{\Cov{\V x\Given\V\theta_{t-1}}^{-1}}^2
  }
\end{align}
where $\mean{\V x}_{|t-1} \bydef \Avg*{\V x\Given\V\theta_{t-1}}$ and
$\Cov{\V x\Given\V\theta_{t-1}}$ are the expectation and the covariance of $\V
x$ knowing $\V\theta_{t-1}$ and where $\Cov{\V z_t}$ is the covariance of the
noise.  This optimization problem is quadratic in $\V x$ and has a closed form
solution:
\begin{align}
  \V x_{|t}
  &= \Paren[\big]{
    \M H_t\T⋅\Cov{\V z_t}^{-1}⋅\M H_t + \Cov{\V x\Given\V\theta_{t-1}}^{-1}
  }^{-1}⋅\Paren[\big]{
    \M H_t\T⋅\Cov{\V z_t}^{-1}⋅\V y_t
    + \Cov{\V x\Given\V\theta_{t-1}}^{-1}⋅\mean{\V x}_{|t-1}
  }\notag\\
  &= \mean{\V x}_{|t-1} + \Paren[\big]{
    \M H_t\T⋅\Cov{\V z_t}^{-1}⋅\M H_t + \Cov{\V x\Given\V\theta_{t-1}}^{-1}
  }^{-1}⋅\M H_t\T⋅\Cov{\V z_t}^{-1}⋅\Paren[\big]{
    \V y_t - \M H_t⋅\mean{\V x}_{|t-1}
  }
\end{align}
by simple algebra.  Now, using a well known matrix
identity\cite{Tarantola-1987-inverse_problem_theory,
  Thiebaut_Tallon-2010-FRiM}, we may write:
\begin{align}
  \V x_{|t}
  &= \mean{\V x}_{|t-1} + \Cov{\V x\Given\V\theta_{t-1}}⋅\M H_t\T⋅\Paren[\big]{
    \Cov{\V z_t} + \M H_t⋅\Cov{\V x\Given\V\theta_{t-1}}⋅\M H_t\T
  }^{-1}⋅\Paren[\big]{
    \V y_t - \M H_t⋅\mean{\V x}_{|t-1}
  }.
\end{align}
Owing to the particular structure of $\V x$ and of $\M H_t$, see
Eqs.~\eqref{eq:wavefront-sequence} and \eqref{eq:H_t}, a few simplifications
can be done:
\begin{align}
  \M H_t⋅\mean{\V x}_{|t-1} &= \M S⋅\mean{\V w}_{t|t-1},\\
  \M H_t⋅\Cov{\V x\Given\V\theta_{t-1}}⋅\M H_t\T &= \M S⋅\Cov{\V w_{t}\Given\V\theta_{t-1}}⋅\M S\T,
\end{align}
with $\mean{\V w}_{t|t-1} \bydef \Avg*{\V w_t\Given\V\theta_{t-1}}$ and
$\Cov{\V w_{t}\Given\V\theta_{t-1}}$ the expectation and covariance of $\V w_t$
knowing $\V\theta_{t-1}$.  Introducing:
\begin{equation}
  \begin{array}{lcccccccr}
    \M Q_{t|t-1} = \bigl( & \cdots & \M 0 & \M 0 & \Cov{\V w_{t}\Given\V\theta_{t-1}}^{-1} & \M 0 & \M 0 & \cdots & \bigr),\\
    & \cdots & t-2 & t-1 & t & t+1 & t+2 & \cdots & \\
  \end{array}
  \label{eq:Q_t}
\end{equation}
it is possible to write $\M H_t = \M S⋅\Cov{\V w_{t}\Given\V\theta_{t-1}}⋅\M Q_{t|t-1}$ and
putting all together, we obtain:
\begin{align}
  \V x_{|t}
  &= \mean{\V x}_{|t-1} +
  \Cov{\V x\Given\V\theta_{t-1}}⋅\M Q_{t|t-1}\T⋅\Cov{\V w_{t}\Given\V\theta_{t-1}}⋅\M S\T⋅\Paren[\big]{
    \Cov{\V z_t} + \M S⋅\Cov{\V w_{t}\Given\V\theta_{t-1}}⋅\M S\T
  }^{-1}⋅\Paren[\big]{
    \V y_t - \M S⋅\mean{\V w}_{t|t-1}
  }\notag\\
  &= \mean{\V x}_{|t-1} +
  \Cov{\V x\Given\V\theta_{t-1}}⋅\M Q_{t|t-1}\T⋅\Paren[\big]{
    \V w_{t|t} - \mean{\V w}_{t|t-1}
  }
\end{align}
where $\V w_{t|t}$ is the MAP estimator of $\V w_t$ knowing $\V\theta_{t}$.
This can be proven using the same rules as before:
\begin{align}
  \V w_{t|t}
  &\bydef \argmax_{\V w_t}\Pr\Paren[\big]{\V w_t \Given \V\theta_{t}} \notag\\
  &= \argmax_{\V w_t}\Pr\Paren[\big]{\V w_t, \V y_t\Given\V\theta_{t-1}} \notag\\
  &= \argmin_{\V w_t}\Brace*{
    \Norm*{\V y_t - \M S⋅\V w_t}_{\Cov{\V z_t}^{-1}}^2
    + \Norm*{\V w_t - \mean{\V w}_{t|t-1}}_{\Cov{\V w_{t}\Given\V\theta_{t-1}}^{-1}}^2
  } \notag\\
  &= \Paren[\big]{
    \M S\T⋅\Cov{\V z_t}^{-1}⋅\M S + \Cov{\V w_{t}\Given\V\theta_{t-1}}^{-1}
  }^{-1}⋅\Paren[\big]{
    \M S\T⋅\Cov{\V z_t}^{-1}⋅\V y_t + \Cov{\V w_{t}\Given\V\theta_{t-1}}^{-1}⋅\mean{\V w}_{t|t-1}
  } \notag\\
  &= \mean{\V w}_{t|t-1} + \Paren[\big]{
    \M S\T⋅\Cov{\V z_t}^{-1}⋅\M S + \Cov{\V w_{t}\Given\V\theta_{t-1}}^{-1}
  }^{-1}⋅\M S\T⋅\Cov{\V z_t}^{-1}⋅\Paren[\big]{
    \V y_t - \M S⋅\mean{\V w}_{t|t-1}
  } \notag\\
  &= \mean{\V w}_{t|t-1} + \Cov{\V w_{t}\Given\V\theta_{t-1}}⋅\M S\T⋅\Paren[\big]{
    \Cov{\V z_t} + \M S⋅\Cov{\V w_{t}\Given\V\theta_{t-1}}⋅\M S\T
  }^{-1}⋅\Paren[\big]{
    \V y_t - \M S⋅\mean{\V w}_{t|t-1}
  }. \quad\blacksquare
\end{align}

The MAP estimator $\V w_{t|t}$ linearly depends on the noise and the wavefront
which are Gaussian variables.  It follows that $\V w_{t|t}$ is also Gaussian,
moreover the MAP estimator is equal to the posterior mean and the associated
conditional covariance $\Cov{\V w_t \Given \V\theta_{t}}$ has a closed form
expression\cite{Tarantola-1987-inverse_problem_theory}:
\begin{align}
  \V w_{t|t} &= \Avg{\V w_t \Given \V\theta_{t}} = \mean{\V w}_{t|t}, \\
  %\Cov{\V w_{t|t}}
  \Cov{\V w_t \Given \V\theta_{t}} &= \Paren[\big]{
    \M S\T⋅\Cov{\V z_t}^{-1}⋅\M S + \Cov{\V w_{t}\Given\V\theta_{t-1}}^{-1}
  }^{-1}.
\end{align}
The MAP estimator of the wavefront sequence is thus also equal to the posterior
mean, that is $\mean{\V x}_{|t} = \V x_{|t}$.% \oops{OK mais un peu rapide ?}

Finally, we have demonstrated the following recurrences (or updating rules):
\begin{align}
  \V w_{t|t} &= \V w_{t|t-1} + \Paren[\big]{
    \M S\T⋅\Cov{\V z_t}^{-1}⋅\M S + \Cov{\V w_{t}\Given\V\theta_{t-1}}^{-1}
  }^{-1}⋅\M S\T⋅\Cov{\V z_t}^{-1}⋅\Paren[\big]{
    \V y_t - \M S⋅\V w_{t|t-1}
  },
  \label{eq:update-w}\\
  %\Cov{\V w_{t|t}}^{-1}
  \Cov{\V w_t \Given \V\theta_{t}}^{-1}
  &= \Cov{\V w_{t}\Given\V\theta_{t-1}}^{-1} + \M S\T⋅\Cov{\V z_t}^{-1}⋅\M S,
  \label{eq:update-cov-w} \\
  \V x_{|t} &=
  \V x_{|t-1} + \M A_{t|t-1}⋅\Paren[\big]{\V w_{t|t} - \V w_{t|t-1}},
  \label{eq:update-x}
\end{align}
where $\M A_{t|t-1} \bydef \Cov{\V x\Given\V\theta_{t-1}}⋅\M Q_{t|t-1}\T$ has
the following block structure:
\begin{equation}
  \M A_{t|t-1}% \bydef \Cov{\V x\Given\V\theta_{t-1}}⋅\M Q_{t|t-1}\T
  = \Paren*{
    \begin{array}{c}
      \vdots \\
      \M B_{t-2,t|t-1}\\
      \M B_{t-1,t|t-1}\\
      \M I\\
      \M B_{t+1,t|t-1}\\
      \M B_{t+2,t|t-1}\\
      \vdots\\
    \end{array}
  }
  \quad \text{with}\quad
  \M B_{t',t|t-1} \bydef \Cov{\V w_{t'},\V w_{t} \Given \V\theta_{t-1}}
  ⋅\Cov{\V w_{t} \Given \V\theta_{t-1}}^{-1},
\end{equation}
where:
\begin{equation}
  \Cov{\V w_{t'},\V w_{t} \Given \V\theta_{t-1}} \bydef \Avg{
    \Paren*{\V w_{t'} - \Avg{\V w_{t'}\Given \V\theta_{t-1}}}⋅
    \Paren*{\V w_{t} - \Avg{\V w_{t}\Given \V\theta_{t-1}}}\T
    \Given\V\theta_{t-1}
  }.
\end{equation}
The matrix $\M B_{t',t|t-1}$ encodes the cross-temporal correlation between
wavefronts at times $t'$ and $t$ knowing information up to time $t-1$.  The
updating rules in Eqs.~\eqref{eq:update-w}, \eqref{eq:update-cov-w} and
\eqref{eq:update-x} assume that the variables (wavefront $\V w_t$ and noise $\V
z_t$ for any frame $t$) are Gaussian, that the noise is centered, that the
noise and the wavefront are independent, and that the noise terms in different
frames are mutually independent.

To close the recurrence, the updating of the wavefront covariance leading to
$\Cov{\V w_{t+1} \Given \V\theta_{t}}$ is required.  This can be obtained from
the updating of the statistics of $\V x$ resulting from the updating of $\V x$
in Eq.~\eqref{eq:update-x}.

Remark that while the first rule in Eq.~\eqref{eq:update-w} is only focused on
the updating of the $t$-th wavefront, Eq.~\eqref{eq:update-cov-w} shows that
the \emph{a posteriori} covariance of the estimated $t$-th wavefront is
reduced, hence that information has been gained, and last rule in
Eq.~\eqref{eq:update-x} propagates this gain in information to all other
frames.  In particular, Eq.~\eqref{eq:update-x} yields the wavefront estimate
$\V w_{t+1|t}$ that will be needed in the next round, that is for frame $t+1$,
of Eqs.~\eqref{eq:update-w} and \eqref{eq:update-x}.  Hence
Eqs.~\eqref{eq:update-w}, \eqref{eq:update-cov-w} and \eqref{eq:update-x},
implement a complete recurrence for updating \emph{complete statistics} of the
unknowns as data arrive.  In fact these equations can be seen as an instance of
the well known \emph{Kalman filter}\cite{Kalman-1960-linear_filtering} for the
case of an AO system.  The variables $\V x$ and the tall matrix $\M A_{t|t-1}$
are the analogous of the so-called \emph{state variables} and \emph{state
  transition matrix} of the Kalman filter.

Taking the $t+δt$-th wavefront in $\V x_{|t}$ yields an updating rule for
$\mean{\V w}_{t+δt|t} = \V w_{t+δt|t}$ the predicted wavefront required in
Eq.~\eqref{eq:optimal-commands}:
\begin{equation}
  \V w_{t+δt|t} = \V w_{t+δt|t-1} + \M B_{t+δt,t|t-1}⋅\Paren{
    \V w_{t|t} - \V w_{t|t-1}
  }
\end{equation}
which require to solve the MAP problem in Eq.~\eqref{eq:update-w}.  How this is
implemented in the actual RTC of the \Themis AO system is detailed in the next
section.

There are several issues for applying the proposed recursion:
\begin{itemize}
\item[(i)] The updating rules do not specify how to initiate the recurrence.
\item[(ii)] The updating rules do not specify how to derive the cross-temporal
  covariances $\Cov{\V w_{t'},\V w_{t} \Given \V\theta_{t-1}}$.  These terms
  may be learned from the available data but this certainly requires some
  assumptions on their structure.
\item[(iii)] In practice, we must work with a truncated sequence, not the
  virtually infinite sequence $\V x$ considered so far.  Owing to the recursive
  structure of the equations, $\V x$ shall be implemented as a sliding temporal
  window of given length.
\item[(iv)] All necessary computations must be carried out faster than the
  rate of the AO system loop.  Typically in less than $1\,\milli\second$ at
  $1\,\kilo\hertz$ for the \Themis AO system.
\end{itemize}

\subsection{Comparison with Conventional AO Systems}

In a simple conventional AO system, the new commands are assumed given by:
\begin{equation}
  \V a_{t+δt} = \V a_{t} - \gamma\,\M G^\ddagger⋅\V d_t
  \label{eq:conventional-commands-update}
\end{equation}
with $\V a_{t}$ the commands at time $t$, $\M G^\ddagger \approx \M G^\dagger$
the WFS data to DM commands matrix which is an approximation of $\M G^\dagger =
\Paren[\big]{\M G\T⋅\M G}^{-1}⋅\M G\T$ the pseudo inverse of the interaction
matrix $\M G$, and $\gamma \in (0,1]$ the control loop gain.  To achieve a
  certain level of regularization, $\M G^\ddagger$ is usually a truncated SVD
  version of $\M G^\dagger$.

Dropping the prediction in the optimal control and directly expressing the
wavefront in the basis of the influence functions of the deformable mirror
amounts to using the commands $\V a_{t+δt} = -\V w_{t|t}$ with $\V w_{t|t}$ the
MAP wavefront estimator given in Eq.~\eqref{eq:update-w}.  Then the commands
are given by:
\begin{align}
  \V a_{t+δt} &= \V a_{t} - \Paren[\big]{
    \M S\T⋅\Cov{\V z_t}^{-1}⋅\M S + \Cov{\V w_{t}\Given\V\theta_{t-1}}^{-1}
  }^{-1}⋅\M S\T⋅\Cov{\V z_t}^{-1}⋅\Paren[\big]{
    \V y_t + \M S⋅\V a_{t}
  } \notag \\
  &= \V a_{t} - \Paren[\big]{
    \M G\T⋅\Cov{\V z_t}^{-1}⋅\M G + \Cov{\V w_{t}\Given\V\theta_{t-1}}^{-1}
  }^{-1}⋅\M G\T⋅\Cov{\V z_t}^{-1}⋅\V d_t,
  \label{eq:regularized-commands-update}
\end{align}
since $\M S = \M G$ in that case and thus $\V y_t + \M S⋅\V a_{t} = \V y_t
+ \M G⋅\V a_{t-1} = \V d_t$.  Even though:

Note the analogies between Eqs.~\eqref{eq:conventional-commands-update} and
\eqref{eq:regularized-commands-update}.  The two approaches are equivalent
provided:
\begin{equation}
  \gamma\,\M G^\ddagger \approx \Paren[\big]{
    \M G\T⋅\Cov{\V z_t}^{-1}⋅\M G + \Cov{\V w_{t}\Given\V\theta_{t-1}}^{-1}
  }^{-1}⋅\M G\T⋅\Cov{\V z_t}^{-1}
\end{equation}
holds to a sufficient precision.

\subsection{The Commands in the Current \Themis AO System}

In the first version of the \Themis AO system, we chose to simplify the
computations so that they remain compatible with a real-time system running on
a conventional CPU and so that they can be controlled with a very small number
of parameters (not all the statistics).  To that end, when solving the MAP
problem in Eq.~\eqref{eq:update-w}, we approximate the wavefront regularization
by:
\begin{equation}
  \Norm*{\V w_t - \V w_{t|t-1}}_{\Cov{\V w_{t}\Given\V\theta_{t-1}}^{-1}}^2
  \approx \mu_t\,\Norm*{\V w_t}^2_{\M C^{-1}} +
  \rho_t\,\Norm*{\V w_t - \V w_{t-1|t-1}}^2
\end{equation}
where $\M C$ is the covariance matrix for a Kolmogorov wavefront with a given
diameter over Fried's parameter $D/r_0$ ratio.  The first term in the above
penalty imposes spatial regularization of the wavefront, the hyper-parameter
$\mu_t \propto (D/r_0)^{-5/3}$ has to be tuned according to the strength of the
turbulence.  The second regularization term imposes the temporal continuity of
the wavefront.  The estimated MAP wavefront at time $t$ is then given by:
\begin{align}
  \V w_{t|t}
  &= \argmin_{\V w_t} \Brace*{
    \Norm*{\V y_t - \M S⋅\V w_t}_{\Cov{\V z_t}^{-1}}^2
    + \mu_t\,\Norm*{\V w_t}^2_{\M C^{-1}}
    + \rho_t\,\Norm*{\V w_t - \V w_{t-1|t-1}}^2
  }\notag\\
  &= \Paren[\big]{
    \M S\T⋅\Cov{\V z_t}^{-1}⋅\M S + \mu_t\,\M C^{-1} + \rho_t\,\M I
  }^{-1}⋅\Paren[\big]{
    \M S\T⋅\Cov{\V z_t}^{-1}⋅\V y_t + \rho_t\,\V w_{t-1|t-1}
  },
  \label{eq:tao-wavefront-estimator}
\end{align}
where the pseudo open-loop data $\V y_t$ are given by
Eq.~\eqref{eq:pol-data-model}.

Unrelated to our simplifying assumptions, an important consequence of the MAP
approach is that while the commands linearly depend on the wavefront sensor
data, the corresponding matrix:
\begin{equation}
  \Paren[\big]{
    \M G\T⋅\Cov{\V z_t}^{-1}⋅\M G + \mu_t\,\M W + \rho_t\,\M I
  }^{-1}⋅\M G\T⋅\Cov{\V z_t}^{-1}
\end{equation}
cannot be computed once for all since it depends on the time.  In other words
the control law changes with time which is an important feature to cope with
variable observing conditions.

Measurements from different sub-pupils can be considered as independent so the
noise covariance matrix $\Cov{\V z_t}$ has a block-diagonal structure with
$2×2$ blocks\footnote{provided horizontal and vertical measured slopes are
  contiguous in the data vector $\V d_t$}.  Each block accounts for the
variances of the horizontal and vertical slopes and for their covariance.  The
measurements $\V d_t$ and their covariances are estimated by the real-time data
processing software of the wavefront sensor (see proceedings by Tallon \emph{et
  al.}\ in this conference\cite{Tallon_et_al-2022-WFS_SPIE}).  Owing to its
block diagonal structure, the noise covariance matrix $\Cov{\V z_t}$ is very
sparse.  The linear response matrix $\M S$ of a Shack-Hartmann wavefront sensor
is also very sparse.  With such sparse matrices, the solution $\V w_{t|t}$ can
be computed iteratively by means of the FRiM
method\cite{Thiebaut_Tallon-2010-FRiM} taking $\M C = \M K⋅\M K\T$ with $\M K$
the \emph{fractal operator} of FRiM.

To further speedup computations, we consider working directly with the
wavefronts represented on the basis of the deformable mirror influence
functions and with no prediction.  This amounts to taking $\M M = \M I$ and $δt
= 0$ in Eq.~\eqref{eq:optimal-commands} and to taking $\M S = \M G$ in all
other equations.  The control commands are then given by:
\begin{align}
  \V a_{t+δt}
  &= \argmin_{\V a} \Brace*{
    \Norm*{\V y_t + \M G⋅\V a}_{\Cov{\V z_t}^{-1}}^2
    +  \mu_t\,\Norm*{\V a}^2_{\M W}
    + \rho_t\,\Norm*{\V a - \V a_{t}}^2
  }\notag\\
  &= \Paren[\big]{
    \M G\T⋅\Cov{\V z_t}^{-1}⋅\M G + \mu_t\,\M W + \rho_t\,\M I
  }^{-1}⋅\Paren[\big]{
    \rho_t\,\V a_{t} - \M G\T⋅\Cov{\V z_t}^{-1}⋅\V y_t
  },
  \label{eq:simplest-commands}
\end{align}
where the pseudo open-loop data $\V y_t$ are given by
Eq.~\eqref{eq:pol-data-model} and where $\M W$ is the inverse of the \emph{a
  priori} spatial covariance of the commands.  The \emph{a priori} precision
matrix $\M W$ can be based on Kolmogorov statistics but for the actuators.

To compare our commands with the ones assumed in simple conventional AO
systems, and given in Eq.~\eqref{eq:conventional-commands-update}, and to the
ones in Eq.~\eqref{eq:regularized-commands-update}, our control commands can be
put in the form of an updating of the previous commands involving the wavefront
sensor data $\V d_t$ instead of the POL data $\V y_t$:
\begin{align}
  \V a_{t+δt}
  &= \V a_{t} -
  \Paren[\big]{\M G\T⋅\Cov{\V z_t}^{-1}⋅\M G + \mu_t\,\M W + \rho_t\,\M I}^{-1}⋅
  \M G\T⋅\Cov{\V z_t}^{-1}⋅\V d_t\notag\\
  &\quad - \mu_t\,
  \Paren[\big]{\M G\T⋅\Cov{\V z_t}^{-1}⋅\M G + \mu_t\,\M W + \rho_t\,\M I}^{-1}⋅
  \M W⋅\V a_{t}.
  \label{eq:simplest-commands-update}
\end{align}

\section{CALIBRATIONS}
\label{sec:calibration}

The optimal control of an AO system relies on the assumed models of the
wavefront sensor, of the deformable mirror, and of the statistics of the
wavefront and of the noise.  The components of the model of the system have to
be carefully calibrated.  We exploit the fact that the AO system includes a
measuring device (the wavefront sensor) and means to modify the system (the
deformable mirror) to perform these calibrations with the AO system itself.

\subsection{Detector Calibration and Image Pre-Processing}
\label{sec:detector}

The calibration of the detector of the wavefront sensor is fully described in
the proceedings by Tallon \emph{et al.}\ in this
conference\cite{Tallon_et_al-2022-WFS_SPIE}.  In a nutshell, this calibration
provides 4 pre-processing parameters per pixel.  Two of these parameters
implement an affine pixel-wise correction to compensate for the non-uniform
bias and overall sensitivity of the detector.  The two other parameters are
needed to estimate the non-uniform precision (reciprocal of the variance) of
the pixels in each pre-processed image.

In the RTC of the \Themis AO system (see Section~\ref{sec:RTC}), a \emph{camera
  server} is continuously acquiring raw images from the wavefront sensor
camera.  As soon as a new raw image is available, the camera server applies the
pre-processing and delivers the pre-processed image and the corresponding pixel
precisions to other processes via shared memory.

\subsection{Live Calibration of the Interaction Matrix}

The interaction matrix $\M G = \M S⋅\M M$ encodes the linear relationship
between the actuator commands and the wavefront sensor data.  In order to
calibrate this matrix, we register data frames of the wavefront sensor with
perturbed commands and given by:
\begin{equation}
  \V d_t = \M S⋅\Paren*{\V w_t + \M M⋅(\V a_t + \varepsilon\,\V u_t)} + \V z_t
  \label{eq:random-perturbation}
\end{equation}
with $\V u_t \in \Reals^{n_\Tag{act}}$ a known vector of random perturbation
and $\varepsilon > 0$ the strength of the perturbation.  The random
perturbations are centered, $\Avg{\V u_t} = \V 0$, mutually independent and
identically distributed (i.i.d.) with a known covariance $\M U = \Cov{\V u_t} =
\Avg{\V u_t⋅\V u_t\T}$.

We then form the following mean cross-product:
\begin{equation}
  \estim{\M V} = \frac{1}{n\,\varepsilon} \sum_{t=1}^{n} \V d_t⋅\V u_t\T
\end{equation}
whose expectation is given by:
\begin{align}
  \Avg[\big]{\estim{\M V}} &= \frac{1}{n\,\varepsilon} \sum_{t=1}^{n} \Brace*{
    \M S ⋅\Paren*{
      \Avg{\V w_t⋅\V u_t\T}
      - \M M⋅\Avg{\V a_t⋅\V u_t\T}
      + \varepsilon\,\M M⋅\Avg{\V u_t⋅\V u_t\T}
    } + \Avg{\V z_t⋅\V u_t\T}
  } \notag\\
  &= \frac{1}{n\,\varepsilon} \sum_{t=1}^{n} \Brace*{
    \M S ⋅\Paren*{
      \Avg{\V w_t}⋅\Avg{\V u_t\T}
      - \M M⋅\Avg{\V a_t}⋅\Avg{\V u_t}\T
      + \varepsilon\,\M M⋅\Avg{\V u_t⋅\V u_t\T}
    } + \Avg{\V z_t}⋅\Avg{\V u_t}\T
  } \notag\\
  &= \frac{1}{n} \, \M S ⋅ \M M ⋅ \sum_{t=1}^{n} \Avg{\V u_t⋅\V u_t\T}
  = \M G ⋅ \M U
  \label{eq:expect-V}
\end{align}
where the first right-and-side (RHS) comes from
Eq.~\eqref{eq:random-perturbation} and from the linearity of the expectation,
the second RHS follows from the independence of $\V u_t$ with all other
variables, and the third RHS from the fact that the perturbations are centered,
\ie $\Avg{\V u_t} = \V 0$, and i.i.d., thus $\Avg{\V u_t⋅\V u_t\T} =
\Cov{\V u_t} = \M U$ for all $t$, and finally using $\M G = \M S ⋅ \M M$.

An \emph{unbiased estimator} of the influence matrix is thus:
\begin{equation}
  \estim{\M G} = \estim{\M V}⋅\M U^{-1}
  \label{eq:calib-G-1}
\end{equation}
with $\M U$ the exact covariance matrix of the perturbations.  For the sake of
simplicity, the perturbations $\V u_t$ may be chosen so that their covariance
matrix is the identity matrix: $\M U = \M I$.

An advantage of this calibration method is that it can be carried out while the
AO system is running.  Indeed, thanks to the fact that the perturbations are
centered and independent to any other variables, the influences of the
turbulent wavefront $\V w_t$, of the actuators commands $\V a_t$, and of the
measurement noise $\V z_t$ cancel on average in $\estim{\M V}$.  The only
restriction is to choose the strength $\varepsilon$ of the perturbation not too
high to not destroy the correction\footnote{since the perturbations are known,
  their impact in the wavefront sensor measurements can be removed provided
  that the influence matrix $\M G$ is known with sufficient accuracy} and yet
high enough to have a measurable incidence in the wavefront sensor data $\V
d_t$.

Since the perturbations are known, it is possible to compute the conditional
expectation of $\estim{\M V}$:
\begin{align}
  \Avg[\big]{\estim{\M V} \Given \Brace{\V u_t}_{t\in1:n}}
  &= \M G ⋅ \estim{\M U} +
  \frac{1}{n\,\varepsilon} \sum_{t=1}^{n} \Brack*{
    \M S ⋅ \Paren*{\Avg{\V w_t} - \M M ⋅ \Avg{\V a_t}}
    + \Avg{\V z_t}
  }⋅\V u_t\T \notag\\
  &= \M G ⋅ \estim{\M U}
  %\text{\quad if $\M S⋅\Paren*{\Avg{\V w_t} - \M M ⋅ \Avg{\V a_t}} + \Avg{\V z_t} = 0$}
  \label{eq:conditional-expectation-of-V}
\end{align}
provided $\M S⋅\Paren*{\Avg{\V w_t} - \M M ⋅ \Avg{\V a_t}} + \Avg{\V z_t} = 0$
holds and with:
\begin{equation}
  \estim{\M U} = \frac{1}{n}\sum_{t=1}^{n}\V u_t⋅\V u_t\T.
\end{equation}
This yields another estimator (which is unbiased conditionally to the knowledge
of the perturbations) for the interaction matrix $\M G$:
\begin{equation}
  \estim{\M G} = \estim{\M V}⋅\estim{\M U}^{-1}.
  \label{eq:calib-G-2}
\end{equation}
From preliminary tests, it seems that Eq.~\eqref{eq:calib-G-2} provides a
better estimator than Eq.~\eqref{eq:calib-G-1}.

In practice, we use a binomial law for the perturbations:
\begin{equation}
  \Brack{\V u_t}_i
  = \begin{cases}
    +1 & \text{with 50\% probability}\\
    -1 & \text{else}
  \end{cases}
\end{equation}
and take $\varepsilon$ equal to a few percents of the dynamics of the
actuators.  This choice yields the covariance matrix $\M U = \M I$.  The
interaction matrix $\M G$ is mostly sparse because the extension of the
influence functions of the deformable mirror are limited.  To get rid of the
noise in the entries of $\M G$ of small amplitude, we only keep the most
significant coefficients of each column of the interaction matrix (that is for
each actuator).  For that, we apply a hard thresholding on the coefficients of
$\estim{\M G}$ as follows:
\begin{equation}
  \estim{G}^\Tag{sparse}_{i,j} =
  \begin{cases}
    \estim{G}_{i,j} & \text{if
        $\sqrt{\estim{G}_{i_\Tag{h}(i),j}^2 + \estim{G}_{i_\Tag{v}(i),j}^2}
        > \max\Paren*{\tau^\Tag{abs}, \tau^\Tag{rel}\,
          \max_{j'}\sqrt{\estim{G}_{i_\Tag{h}(i),j'}^2 +
            \estim{G}_{i_\Tag{v}(i),j'}^2}}$}\\
      0 & \text{else}
  \end{cases}
\end{equation}
to where $i_\Tag{h}(i)$ and $i_\Tag{v}(i)$ yield the index of the wavefront
measurement corresponding to the slopes respectively along the horizontal and
vertical axes for the $i$-th measurement and where $\tau^\Tag{abs} ≥ 0$ and
$\tau^\Tag{rel} ≥ 0$ are chosen absolute and relative threshold levels.
Assuming 1-based indexing and that the 2 slopes measured for a given sub-image
are contiguous in the data, then $i_\Tag{h}(i) = 2\,\Floor{(i-1)/2} + 1$ and
$i_\Tag{v}(i) = i_\Tag{h}(i) + 1$.

\section{Real-Time Control}
\label{sec:RTC}

\begin{figure}
  \centering
  \includegraphics[width=\textwidth]{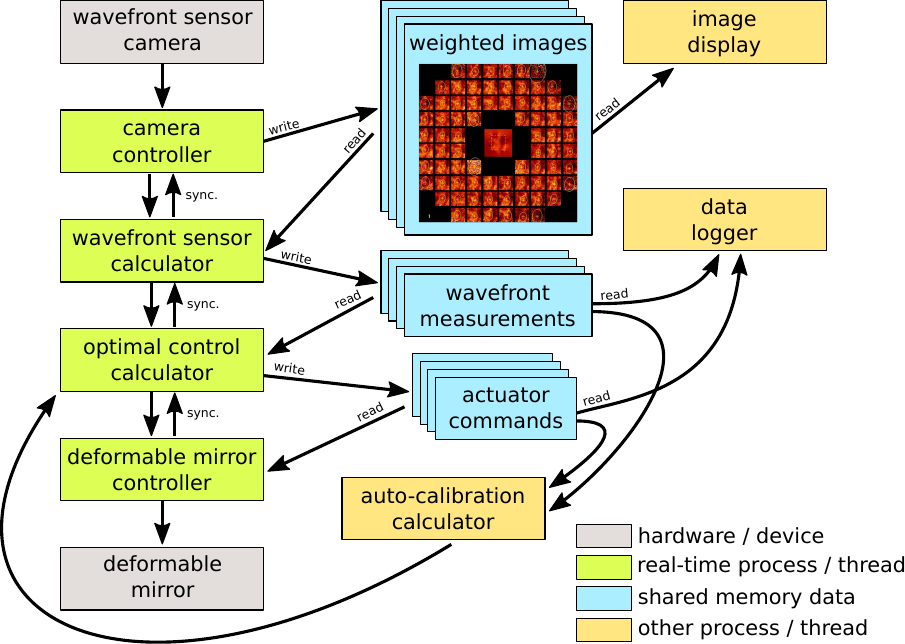}
  \caption{The RTC framework of \Themis.}
  \label{fig:RTC}
\end{figure}

The real time control (RTC) software of the \Themis AO system runs on a Linux
workstation with a 4 core i7-4790K CPU at $4.2\,\GHz$.  The Linux kernel is a
\emph{low-latency} version.  The RTC of the \Themis AO system consists in
multiple processes which collaborate to execute the various tasks (see
Fig.~\ref{fig:RTC}).  The time critical processes (wavefront sensor camera
sever, wavefront sensor calculator, control command calculator, and deformable
mirror server) run with a real-time priority to avoid jitters and latency in
their execution.  Other non-critical processes (visualization, telemetry
collector, statistics calculator, \etc) run with normal priorities.  All theses
processes use a common framework called TAO (which is the acronym for \emph{a
  Toolkit for Adaptive Optics systems}) which is a set of libraries written in
C and bindings for different languages (currently C,
Julia\cite{Bezanson_et_al-2017-Julia}, Yorick, and Unix shell).  TAO is meant
to be portable and flexible.  Several models of devices (cameras, deformable
mirrors, LCD phase screens, and step motor controllers) have been interfaced in
TAO.  Most of the software is publicly
available\footnote{\url{https://git-cral.univ-lyon1.fr/tao}}.

In the TAO framework, devices (the wavefront camera and the deformable mirror)
are managed by servers:
\begin{itemize}
\item The server owning the \textbf{wavefront camera} continuously acquires
  images as quickly as possible, applies image pre-processing (affine
  correction and estimation of the pixel-wise precision) and delivers the
  resulting \emph{weighted images} via shared memory to other processes.
  Shared read/write locks and condition variables are used to notify the other
  processes that a new image is available.

\item The server owning the \textbf{deformable mirror} (DM) may receive
  actuators commands from any other process and apply these commands to the DM
  accounting for given reference commands preset so as to compensate for non
  common path aberrations (NCPA).  The server stores in shared memory a cyclic
  history of the commands effectively applied to the DM (taking into account
  the reference and limitations of the DM such as minimal and maximal command
  levels).  This history is updated immediately after applying the commands and
  other processes are notified so that they can figure out exactly which
  commands have been sent.  This is needed, in particular, to compute the
  pseudo open-loop data.
\end{itemize}
The servers share information and receive commands from other processes via
resources stored in shared memory and whose access is controlled by shared
read/write locks and condition variables.  With critical processes ran with
real-time priority, the latency between a condition being notified and the
process waiting for this condition to be awaken is small and quite stable
(about $6\pm1\,\usec$).  Latency of the same order have been measured for other
RTC software like \textsc{Cacao}\cite{Guyon_et_al-2020-CACAO_SPIE} relying on
semaphores for process synchronization.  We chose read/write locks and
conditions variables so as to be the most flexible regarding which clients are
allowed to connect to which resources and we rely on real-time priorities and
the Linux process scheduler to dispatch the available CPU power.  Compared to
semaphores, read/write locks (like mutexes) may result in dead-locks if
incorrectly used.

There are several advantages in the splitting of the tasks among several
processes and in the sharing of information in shared memory:
\begin{itemize}
\item There are no latency due to transferring/copying of data.
\item The available CPU power is better exploited and the latency reduced
  (in spite of the small delays induced by process synchronization) because
  a process can deliver the result of its work to others as soon as possible
  and then executes post-processing tasks.
\item Non-sequential coding is more appropriate for implementing an AO system
  RTC and it turns out that the resulting code is considerably simpler than an
  equivalent version implemented with sequential coding.
\item Once a device server is launched, any other process have access to the
  device and there is no restrictions on the number of clients accessing a
  given resources.
\end{itemize}

\begin{figure}
  \centering
  \includegraphics[width=150mm]{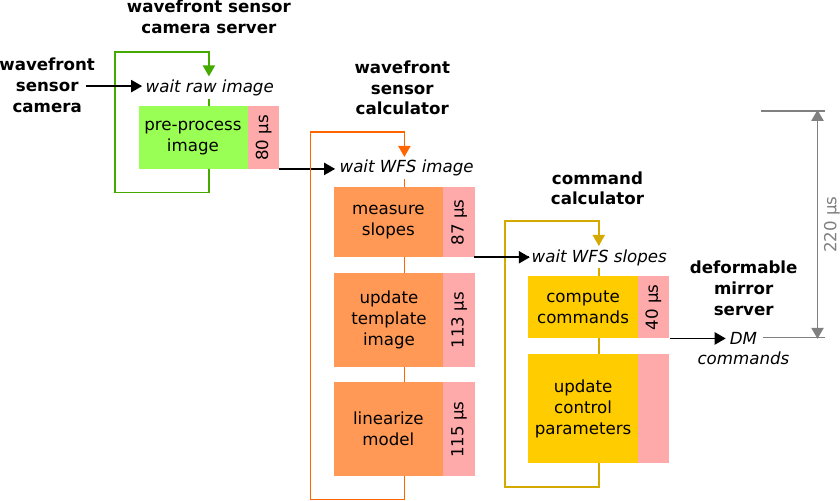}
  \caption{Scheduling of real-time tasks in \Themis AO control.  The RTC is
    able to deliver deformable mirros commands with a delay of
    $\simeq220\,\usec$ from the acquisition of the raw wavefront sensor image}
  \label{fig:tao-rtc-schedule}
\end{figure}

Accounting for the synchronization of tasks, \Themis AO RTC is able to send
commands to the deformable mirror with a delay of $\simeq220\,\usec$ after
receiving the raw image from the wavefront sensor camera (see
Fig.~\ref{fig:tao-rtc-schedule}).  The RTC was tested with a loop frequency up
to 1.2\,kHz.

\section{First closing of the loop and first images}
\label{sec:results}

\begin{figure}
  \centering
  \includegraphics[height=55mm]{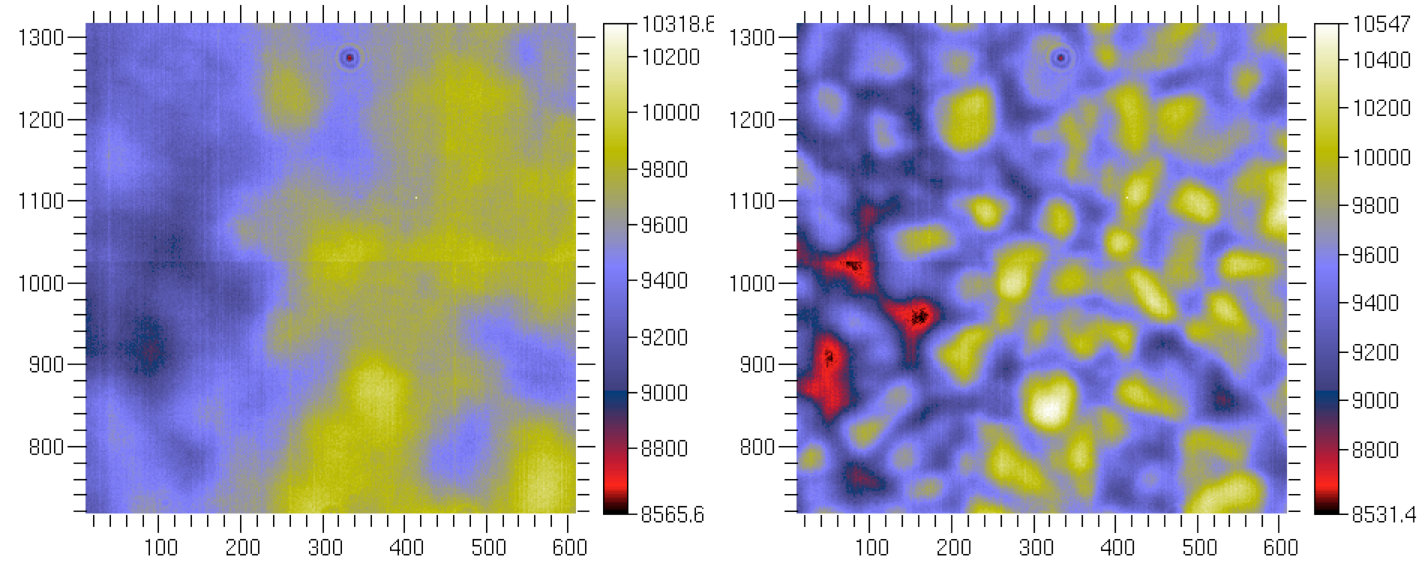}
  \caption{First closing of the loop (8 December 2020).  Images of a 14.6''
    field of view (FOV) of the Sun in a $H_\alpha$ filter and centered at the
    FOV of the wavefront sensor.  Left: open loop.  Right: closed loop at
    1\,kHz.  Temporal smoothing of 0.5\,s, Fried's parameter estimated to be
    $r_0\simeq5\,\centi\meter$.}
  \label{fig:solar-granulation-1}
\end{figure}

\begin{figure}
  \centering
  \includegraphics[height=55mm]{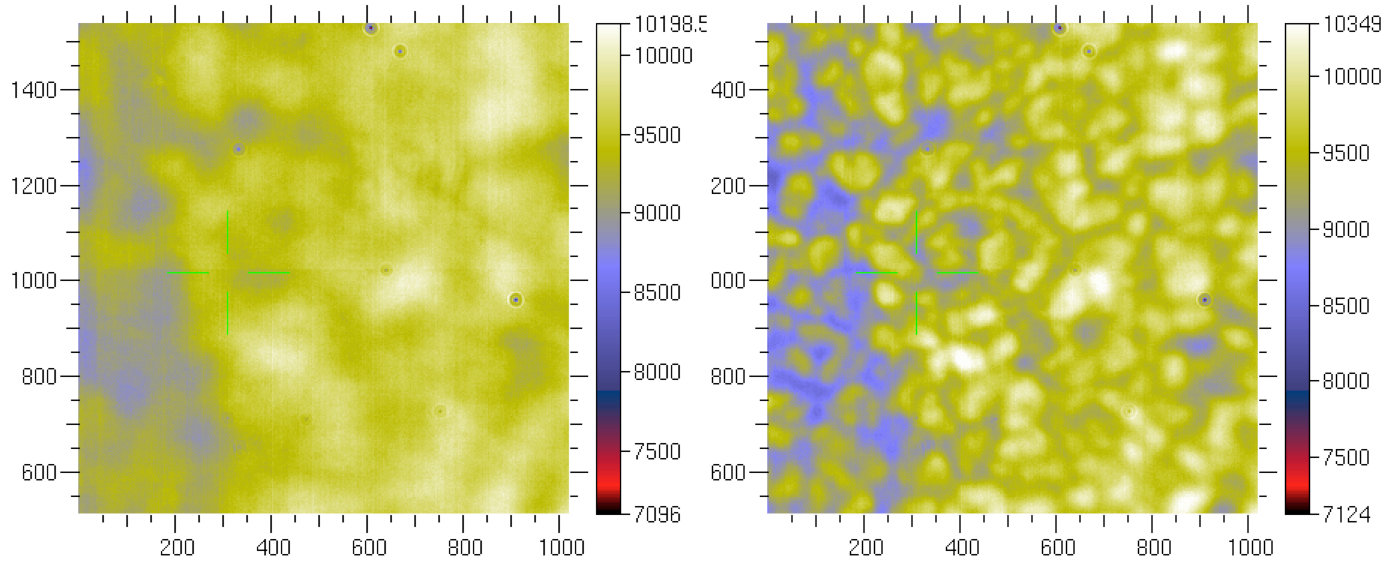}
  \caption{Same as Fig.~\ref{fig:solar-granulation-1} but for a larger FOV of
    25'' and in worse turbulence conditions: Fried's parameter estimated to be
    $r_0\simeq3\,\centi\meter$.  The green cross indicates the center of the
    wavefront sensor 10'' FOV.}
  \label{fig:solar-granulation-2}
\end{figure}

\begin{figure}
  \centering
  \includegraphics[height=55mm]{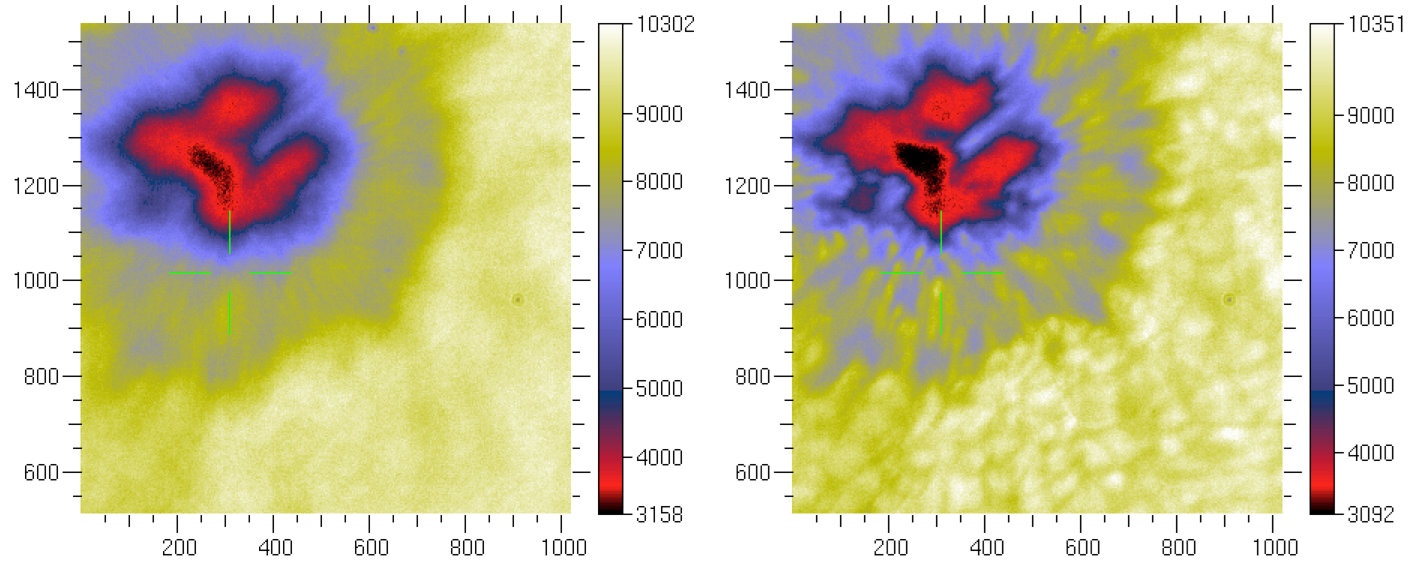}
  \caption{Sun spot observed on 8th of December 2020.  Left: open loop.  Right:
    closed loop at 1\,kHz.  25'' FOV.  The green cross indicates the center of
    the wavefront sensor 10'' FOV.}
  \label{fig:sun-spot}
\end{figure}

\begin{figure}
  \centering
  \includegraphics[width=\textwidth]{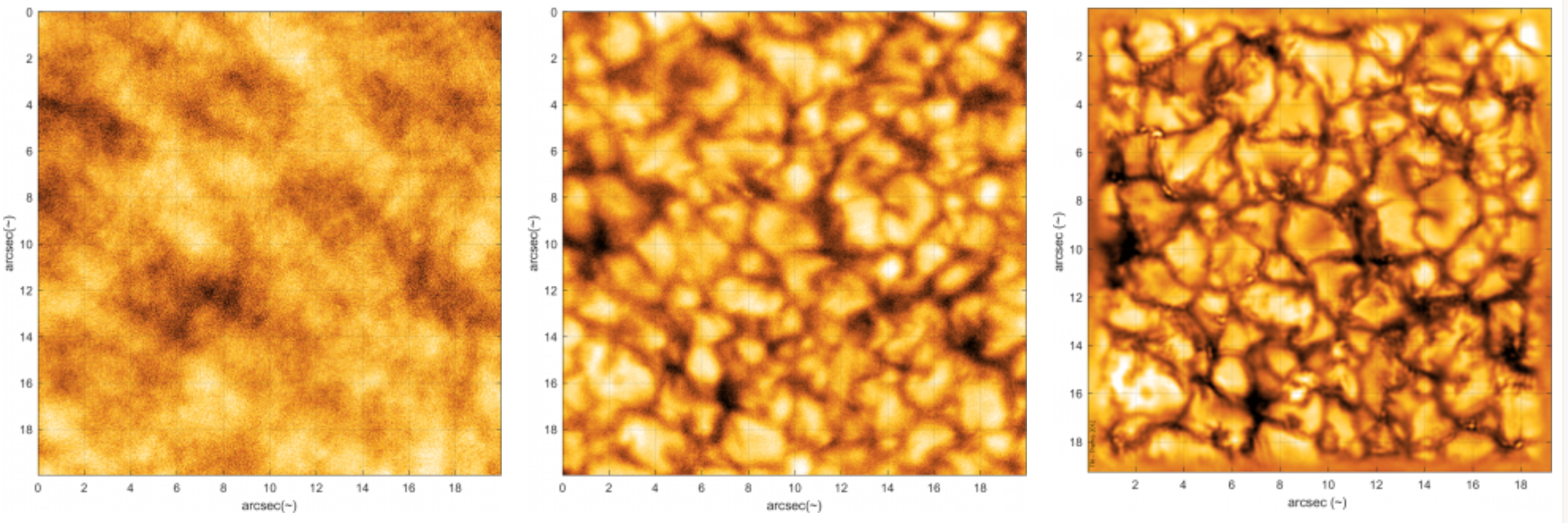}
  \caption{Solar granulation observed with \Themis AO system.  Left: open loop,
    granulation contrast: 1.7\,\%.  Middle: closed loop, granulation contrast:
    4.2\,\%.  Right: result of post-processing of 100 images in closed loop
    (Knox-Thompson), granulation contrast: 9.6\,\%.  Fried's parameter
    estimated to be $r_0\simeq3-4\,\centi\meter$.}
  \label{fig:solar-granulation-with-postprocessing}
\end{figure}

\begin{figure}
  \centering
  \includegraphics[width=145mm]{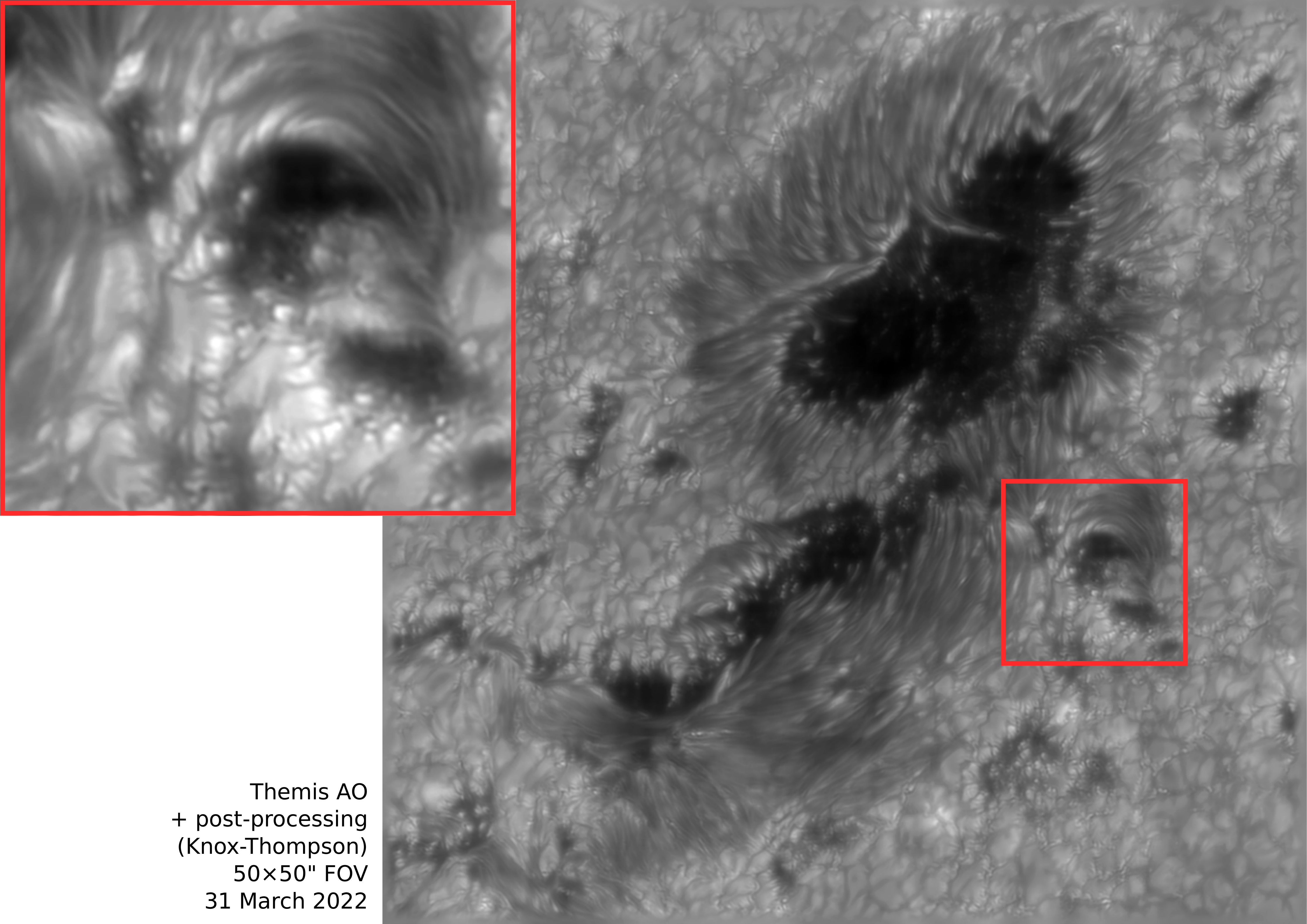}
  \caption{Sun spots with \Themis AO system and post-processed of images in
    closed loop.}
  \label{fig:sun-spots-1}
\end{figure}

\begin{figure}
  \centering
  \includegraphics[width=145mm]{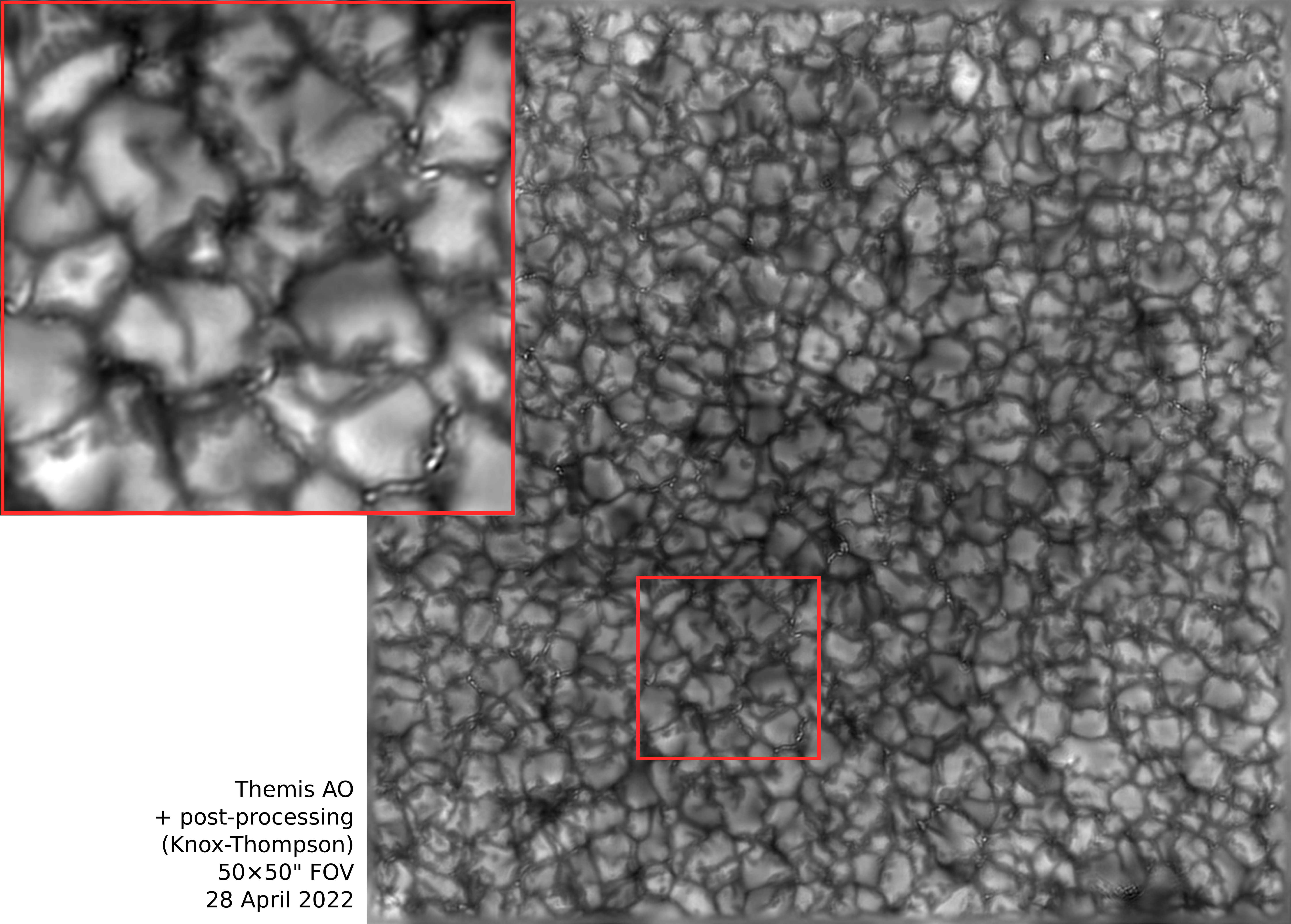}
  \caption{Solar granulation with \Themis AO system and post-processed of
    images in closed loop.}
  \label{fig:solar-granulation-3}
\end{figure}

We closed the loop for the first time on the 8th of December 2020.
Figures~\ref{fig:solar-granulation-1}--\ref{fig:sun-spot} show some images of
the Sun in open and closed loop at that time\footnote{For more details and
  movies, see
  \url{https://git-cral.univ-lyon1.fr/isabelle.tallon-bosc/tao-news/} and
  \url{http://161.72.34.10/dokuwiki/doku.php?id=themis_ao_first_light}.}.
Figure~\ref{fig:solar-granulation-with-postprocessing} shows the benefits of
the AO system but also of the post-processing of images acquired in closed loop
(here by a Knox-Thompson method).  Figures~\ref{fig:solar-granulation-3} and
\ref{fig:sun-spots-1} show the results obtained by post-processing images
recently acquired in closed loop with the \Themis AO system in March and April
2022.  These images demonstrate the great potential of this instrument (the
fine grains seen between the granules are caused by small scale structures of
the magnetic field and have a size of about 0.2'' which is close to the
diffraction limit of \Themis).

\acknowledgments % equivalent to \section*{ACKNOWLEDGMENTS}

The project has been funded through the FP7 program INFRA-2012-1.1.26, SOLARNET
grant agreement n°~312495 of the European Commission.  It has also obtained
support from the ``\emph{Action Spécifique Haute Résolution Angulaire}''
(ASHRA) of CNRS-INSU co-funded by CNES and the Actions Incitatives of the
``\emph{Centre de Recherches Astrophysiques de Lyon}'' (CRAL).

% References
\bibliography{biblio} % bibliography data in report.bib
\bibliographystyle{spiebib} % makes bibtex use spiebib.bst

\end{document}

%% file: new-macros.tex
\usepackage{amsmath,amsfonts,amssymb}
\usepackage{mathtools}

\usepackage[utf8]{inputenc}

 \DeclareUnicodeCharacter{20AC}{\euro} % Euro sign
 \DeclareUnicodeCharacter{2070}{\ensuremath{^0}} % ⁰
 \DeclareUnicodeCharacter{00B9}{\ensuremath{^1}} % ¹
 \DeclareUnicodeCharacter{00B2}{\ensuremath{^2}} % ²
 \DeclareUnicodeCharacter{00B3}{\ensuremath{^3}} % ³
 \DeclareUnicodeCharacter{2074}{\ensuremath{^4}} % ⁴
 \DeclareUnicodeCharacter{2075}{\ensuremath{^5}} % ⁵
 \DeclareUnicodeCharacter{2076}{\ensuremath{^6}} % ⁶
 \DeclareUnicodeCharacter{2077}{\ensuremath{^7}} % ⁷
 \DeclareUnicodeCharacter{2078}{\ensuremath{^8}} % ⁸
 \DeclareUnicodeCharacter{2079}{\ensuremath{^9}} % ⁹

 \DeclareUnicodeCharacter{0391}{\ensuremath{\Alpha}}     % (Α) greek Alpha
 \DeclareUnicodeCharacter{0392}{\ensuremath{\Beta}}      % (Β) greek Beta
 \DeclareUnicodeCharacter{0393}{\ensuremath{\Gamma}}     % (Γ) greek Gamma
 \DeclareUnicodeCharacter{0394}{\ensuremath{\Delta}}     % (Δ) greek Delta
 \DeclareUnicodeCharacter{0395}{\ensuremath{\Epsilon}}   % (Ε) greek Epsilon
 \DeclareUnicodeCharacter{0396}{\ensuremath{\Zeta}}      % (Ζ) greek Zeta
 \DeclareUnicodeCharacter{0397}{\ensuremath{\Eta}}       % (Η) greek Eta
 \DeclareUnicodeCharacter{0398}{\ensuremath{\Theta}}     % (Θ) greek Theta
 \DeclareUnicodeCharacter{0399}{\ensuremath{\Iota}}      % (Ι) greek Iota
 \DeclareUnicodeCharacter{039A}{\ensuremath{\Kappa}}     % (Κ) greek Kappa
 \DeclareUnicodeCharacter{039B}{\ensuremath{\Lambda}}    % (Λ) greek Lambda
 \DeclareUnicodeCharacter{039C}{\ensuremath{\Mu}}        % (Μ) greek Mu
 \DeclareUnicodeCharacter{039D}{\ensuremath{\Nu}}        % (Ν) greek Nu
 \DeclareUnicodeCharacter{039E}{\ensuremath{\Xi}}        % (Ξ) greek Xi
 \DeclareUnicodeCharacter{039F}{\ensuremath{\Omicron}}   % (Ο) greek Omicron
 \DeclareUnicodeCharacter{03A0}{\ensuremath{\Pi}}        % (Π) greek Pi
 \DeclareUnicodeCharacter{03A1}{\ensuremath{\Rho}}       % (Ρ) greek Rho
 \DeclareUnicodeCharacter{03A3}{\ensuremath{\Sigma}}     % (Σ) greek Sigma
 \DeclareUnicodeCharacter{03A4}{\ensuremath{\Tau}}       % (Τ) greek Tau
 \DeclareUnicodeCharacter{03A5}{\ensuremath{\Upsilon}}   % (Υ) greek Upsilon
 \DeclareUnicodeCharacter{03A6}{\ensuremath{\Phi}}       % (Φ) greek Phi
 \DeclareUnicodeCharacter{03A7}{\ensuremath{\Chi}}       % (Χ) greek Chi
 \DeclareUnicodeCharacter{03A8}{\ensuremath{\Psi}}       % (Ψ) greek Psi
 \DeclareUnicodeCharacter{03A9}{\ensuremath{\Omega}}     % (Ω) greek Omega

 \DeclareUnicodeCharacter{03B1}{\ensuremath{\alpha}}     % (α) greek alpha
 \DeclareUnicodeCharacter{03B2}{\ensuremath{\beta}}      % (β) greek beta
 \DeclareUnicodeCharacter{03B3}{\ensuremath{\gamma}}     % (γ) greek gamma
 \DeclareUnicodeCharacter{03B4}{\ensuremath{\delta}}     % (δ) greek delta
 \DeclareUnicodeCharacter{03B5}{\ensuremath{\varepsilon}}% (ε) greek epsilon
 \DeclareUnicodeCharacter{03B6}{\ensuremath{\zeta}}      % (ζ) greek zeta
 \DeclareUnicodeCharacter{03B7}{\ensuremath{\eta}}       % (η) greek eta
 \DeclareUnicodeCharacter{03B8}{\ensuremath{\theta}}     % (θ) greek theta
 \DeclareUnicodeCharacter{03B9}{\ensuremath{\iota}}      % (ι) greek iota
 \DeclareUnicodeCharacter{03BA}{\ensuremath{\kappa}}     % (κ) greek kappa
 \DeclareUnicodeCharacter{03BB}{\ensuremath{\lambda}}    % (λ) greek lambda
 \DeclareUnicodeCharacter{03BC}{\ensuremath{\mu}}        % (μ) greek mu
 \DeclareUnicodeCharacter{03BD}{\ensuremath{\nu}}        % (ν) greek nu
 \DeclareUnicodeCharacter{03BE}{\ensuremath{\xi}}        % (ξ) greek xi
 \DeclareUnicodeCharacter{03BF}{\ensuremath{\omicron}}   % (ο) greek omicron
 \DeclareUnicodeCharacter{03C0}{\ensuremath{\pi}}        % (π) greek pi
 \DeclareUnicodeCharacter{03C1}{\ensuremath{\rho}}       % (ρ) greek rho
 \DeclareUnicodeCharacter{03C2}{\ensuremath{\varsigma}}  % (ς) greek stigma
 \DeclareUnicodeCharacter{03C3}{\ensuremath{\sigma}}     % (σ) greek sigma
 \DeclareUnicodeCharacter{03C4}{\ensuremath{\tau}}       % (τ) greek tau
 \DeclareUnicodeCharacter{03C5}{\ensuremath{\upsilon}}   % (υ) greek upsilon
 \DeclareUnicodeCharacter{03C6}{\ensuremath{\varphi}}    % (φ) greek phi
 \DeclareUnicodeCharacter{03C7}{\ensuremath{\chi}}       % (χ) greek chi
 \DeclareUnicodeCharacter{03C8}{\ensuremath{\psi}}       % (ψ) greek psi
 \DeclareUnicodeCharacter{03C9}{\ensuremath{\omega}}     % (ω) greek omega

 \DeclareUnicodeCharacter{1D538}{\ensuremath{\mathbb{A}}} % (𝔸)
 \DeclareUnicodeCharacter{1D539}{\ensuremath{\mathbb{B}}} % (𝔹)
 \DeclareUnicodeCharacter{2102}{\ensuremath{\mathbb{C}}}  % (ℂ)
 \DeclareUnicodeCharacter{1D53B}{\ensuremath{\mathbb{D}}} % (𝔻)
 \DeclareUnicodeCharacter{1D53C}{\ensuremath{\mathbb{E}}} % (𝔼)
 \DeclareUnicodeCharacter{1D53D}{\ensuremath{\mathbb{F}}} % (𝔽)
 \DeclareUnicodeCharacter{1D53E}{\ensuremath{\mathbb{G}}} % (𝔾)
 \DeclareUnicodeCharacter{210D}{\ensuremath{\mathbb{H}}}  % (ℍ)
 \DeclareUnicodeCharacter{1D540}{\ensuremath{\mathbb{I}}} % (𝕀)
 \DeclareUnicodeCharacter{1D541}{\ensuremath{\mathbb{J}}} % (𝕁)
 \DeclareUnicodeCharacter{1D542}{\ensuremath{\mathbb{K}}} % (𝕂)
 \DeclareUnicodeCharacter{1D543}{\ensuremath{\mathbb{L}}} % (𝕃)
 \DeclareUnicodeCharacter{1D544}{\ensuremath{\mathbb{M}}} % (𝕄)
 \DeclareUnicodeCharacter{2115}{\ensuremath{\mathbb{N}}}  % (ℕ)
 \DeclareUnicodeCharacter{1D546}{\ensuremath{\mathbb{O}}} % (𝕆)
 \DeclareUnicodeCharacter{2119}{\ensuremath{\mathbb{P}}}  % (ℙ)
 \DeclareUnicodeCharacter{211A}{\ensuremath{\mathbb{Q}}}  % (ℚ)
 \DeclareUnicodeCharacter{211D}{\ensuremath{\mathbb{R}}}  % (ℝ)
 \DeclareUnicodeCharacter{1D54A}{\ensuremath{\mathbb{S}}} % (𝕊)
 \DeclareUnicodeCharacter{1D54B}{\ensuremath{\mathbb{T}}} % (𝕋)
 \DeclareUnicodeCharacter{1D54C}{\ensuremath{\mathbb{U}}} % (𝕌)
 \DeclareUnicodeCharacter{1D54D}{\ensuremath{\mathbb{V}}} % (𝕍)
 \DeclareUnicodeCharacter{1D54E}{\ensuremath{\mathbb{W}}} % (𝕎)
 \DeclareUnicodeCharacter{1D54F}{\ensuremath{\mathbb{X}}} % (𝕏)
 \DeclareUnicodeCharacter{1D550}{\ensuremath{\mathbb{Y}}} % (𝕐)
 \DeclareUnicodeCharacter{2124}{\ensuremath{\mathbb{Z}}}  % (ℤ)

 \DeclareUnicodeCharacter{00B0}{\ensuremath{^{\circ}}} % °
 \DeclareUnicodeCharacter{00B1}{\ensuremath{\pm}} % plus-minus sign
 \DeclareUnicodeCharacter{00B5}{\ensuremath{+\upmu+}} % (µ) micro sign
 \DeclareUnicodeCharacter{00B7}{\ensuremath{\cdot}} % (⋅) centered dot
 \DeclareUnicodeCharacter{00D7}{\ensuremath{\times}} % (×) multiplication sign
 \DeclareUnicodeCharacter{00F7}{\ensuremath{\div}} % (÷) division sign
 \DeclareUnicodeCharacter{2026}{\ensuremath{\ldots}} % (…) ellipsis
 \DeclareUnicodeCharacter{2113}{\ensuremath{\ell}} % (ℓ)
 \DeclareUnicodeCharacter{2200}{\ensuremath{\forall}} % (∀) for all
 \DeclareUnicodeCharacter{2202}{\ensuremath{\partial}} % (∂) partial differential
 \DeclareUnicodeCharacter{2205}{\ensuremath{\varnothing}} % (∅) empty set
 \DeclareUnicodeCharacter{2207}{\ensuremath{\nabla}} % (∇) nabla, gradient
 \DeclareUnicodeCharacter{2208}{\ensuremath{\in}}     % (∈) element of
 \DeclareUnicodeCharacter{2209}{\ensuremath{\not\in}} % (∉) not an element of
 \DeclareUnicodeCharacter{22A4}{\ensuremath{\top}} % (⊤) down tack, top
 \DeclareUnicodeCharacter{2218}{\ensuremath{\circ}} % (∘) composition of functions
 \DeclareUnicodeCharacter{221D}{\ensuremath{\propto}} % (∝) proportional to
 \DeclareUnicodeCharacter{221E}{\ensuremath{\infty}} % (∞) infinity
 \DeclareUnicodeCharacter{2220}{\ensuremath{\angle}} % (∠) angle
 \DeclareUnicodeCharacter{2248}{\ensuremath{\approx}} % (≈)
 \DeclareUnicodeCharacter{2260}{\ensuremath{\not=}}
 \DeclareUnicodeCharacter{2261}{\ensuremath{\equiv}} % ()
 \DeclareUnicodeCharacter{2264}{\ensuremath{\le}}    % ≤
 \DeclareUnicodeCharacter{2265}{\ensuremath{\ge}}    % ≥
 \DeclareUnicodeCharacter{2297}{\ensuremath{\otimes}} % ()
 \DeclareUnicodeCharacter{2299}{\ensuremath{\odot}} % ()
 \DeclareUnicodeCharacter{22A4}{\ensuremath{\top}} % ()
 \DeclareUnicodeCharacter{22C5}{\ensuremath{\cdot}} % (⋅) dot operator
 \DeclareUnicodeCharacter{2715}{\ensuremath{\times}}
 \DeclareUnicodeCharacter{27FA}{\ensuremath{\Longleftrightarrow}} % ⟺
 \DeclareUnicodeCharacter{2A2F}{\ensuremath{\times}} % cross product

 \DeclareUnicodeCharacter{2190}{\ensuremath{\leftarrow}}
 \DeclareUnicodeCharacter{2191}{\ensuremath{\uparrow}}
 \DeclareUnicodeCharacter{2192}{\ensuremath{\rightarrow}} % (→) rightwards arrow
 \DeclareUnicodeCharacter{2193}{\ensuremath{\downarrow}}
 \DeclareUnicodeCharacter{21A6}{\ensuremath{\mapsto}} % (↦)
 \DeclareUnicodeCharacter{21D2}{\ensuremath{\Rightarrow}} % (⇒)

\usepackage{graphicx}
\graphicspath{{figs/}}

\usepackage{lmodern}

\DeclareMathAlphabet{\mymathbb}{U}{BOONDOX-ds}{m}{n}

\usepackage[bb=ams]{mathalfa}
\usepackage[ruled,vlined]{algorithm2e}

\SetKwComment{Comment}{}{}

\usepackage{xspace}

\newcommand*{\etc}{\emph{etc.}\xspace}

\newcommand*{\ie}{\emph{i.e.}\xspace}

\usepackage[squaren,Gray]{SIunits}
\addunit{\flops}{flops}

\DeclareSymbolFont{bbold}{U}{bbold}{m}{n}
\DeclareSymbolFontAlphabet{\mathbbold}{bbold}
\RequirePackage{textcomp}  % required for special glyphs

\DeclarePairedDelimiterX{\Paren}[1]{(}{)}{#1}
\DeclarePairedDelimiterX{\Brace}[1]{\{}{\}}{#1}
\DeclarePairedDelimiterX{\Brack}[1]{[}{]}{#1}
\DeclarePairedDelimiterX{\Abs}[1]{\rvert}{\lvert}{#1}
\DeclarePairedDelimiterX{\Norm}[1]{\lVert}{\rVert}{#1}
\DeclarePairedDelimiterX{\Avg}[1]{\langle}{\rangle}{#1}
\DeclarePairedDelimiterX{\Round}[1]{\lfloor}{\rceil}{#1}
\DeclarePairedDelimiterX{\Floor}[1]{\lfloor}{\rfloor}{#1}
\DeclarePairedDelimiterX{\Ceil}[1]{\lceil}{\rceil}{#1}
\DeclarePairedDelimiterX{\Inner}[1]{\langle}{\rangle}{#1}
\DeclarePairedDelimiterX{\IntRange}[1]{\llbracket}{\rrbracket}{#1}
\DeclarePairedDelimiterX{\Group}[1]{\lgroup}{\rgroup}{#1}
\DeclarePairedDelimiterXPP{\List}[3]{}{\{}{\}}{_{#2,\ldots,#3}}{#1}

\newcommand*{\delimsize}{}
\newcommand*{\Given}{\,\delimsize\vert\,} % autoscale to surrounding version
\DeclareMathOperator*{\argmin}{arg\,min}
\DeclareMathOperator*{\argmax}{arg\,max}

\DeclarePairedDelimiterXPP{\Var}[1]{\mathrm{Var}}(){}{#1}
\DeclarePairedDelimiterXPP{\Cov}[1]{\mathrm{Cov}}(){}{#1}
\DeclarePairedDelimiterXPP{\Expect}[1]{\mathrm{E}}(){}{#1}
\DeclarePairedDelimiterXPP{\LogPr}[1]{\ell}(){}{#1}

\newcommand*{\Set}[1]{\mathbb{#1}}
\newcommand*{\Reals}{\Set{R}}

\newcommand*{\Tag}[1]{\mathrm{#1}}

\newcommand*{\bydef}{\stackrel{{\scriptscriptstyle \text{def}}}{=}}

\newcommand*{\V}[1]{\boldsymbol{#1}}   % a vector
\newcommand*{\M}[1]{\boldsymbol{#1}}   % an operator (a matrix)

\newcommand*{\Adj}{\top}
\newcommand*{\T}{^{\Adj}}

\renewcommand*{\cdot}{\mathord{\,\mathchar"2201\,}}
  \def\clap#1{\hbox to 0pt{\hss#1\hss}}